\documentclass{aa}
\usepackage[varg]{txfonts}
\usepackage{graphicx}
\usepackage{natbib}
\usepackage{xcolor}
\begin{document}

\title{Wind accretion in Cygnus X-1}
\author{E. Meyer-Hofmeister\inst{1}
  \and B.F. Liu\inst{2,3}
  \and E. Qiao\inst{2,3}
  \and Ronald E. Taam\inst{4}
}
\offprints{Emmi Meyer-Hofmeister, \email{emm@mpa-garching.mpg.de}}
\institute
    {Max-Planck-Institut f\"ur Astrophysik, Karl-
      Schwarzschildstr.~1, D-85740 Garching, Germany
    \and
    Key Laboratory of Space Astronomy and Technology, National
    Astronomical Observatories, Chinese Academy of Sciences, Beijing
    100012, China
    \and
    School of Astronomy and Space Science, University of Chinese
    Academy of Sciences, 19A Yuquan Road, Beijing 100049, China
    \and 
    Center for Interdisciplinary Exploration and Research in
    Astrophysics
    (CIERA), Department of Physics and Astronomy, Northwestern University,
    2145 Sheridan Road, Evanston, IL 60208, USA
    }
\date{Received: / Accepted:}

\abstract
  {Cygnus X-1 is a black hole X-ray binary system in which the black hole captures and accretes gas from 
     the strong stellar wind emitted by its supergiant O9.7 companion
     star. The irradiation of the 
     supergiant star essentially determines the flow properties of the stellar wind and the X-ray 
     luminosity from the system.  The results of three-dimensional hydrodynamical simulations of wind-fed X-ray binary systems reported in recent work  reveal that the ionizing feedback of the X-ray 
     irradiation leads to the existence of two stable states with either a soft or a hard spectrum.}
  {We discuss the observed radiation of Cygnus X-1 in the soft and hard state in the context of mass flow in 
     the corona and disk, as predicted by the recent application of a condensation model.}
  {The rates of gas condensation from the corona to the disk for Cygnus X-1 are determined, and the 
    spectra of the hard and soft radiation are computed. The theoretical results are compared with the 
    MAXI observations of Cygnus X-1 from 2009 to 2018. In particular, we evaluate the 
    hardness-intensity diagrams (HIDs) for its ten episodes of soft and hard states which show that Cygnus X-1
    is distinct in its spectral changes as compared to those found in the HIDs of low-mass X-ray binaries.}
  {The theoretically derived values of photon counts and hardness are in approximate agreement 
    with the observed data in the HID. However, the scatter in the
    diagram is not reproduced. Improved agreement 
    could result from variations in the viscosity associated with clumping in the stellar wind and 
    corresponding changes of the magnetic fields in the disk. The observed dipping events in the 
    hard state may also contribute to the scatter and to a harder spectrum than predicted by the model.}
    {}

\maketitle

\section{Introduction}
Cygnus X-1 is one of the most frequently observed high-mass X-ray binaries (HMXB) having been studied with 
instruments on ground-based and space-based telescopes.
The source consists of a compact object, a black hole, and a blue supergiant star, HDE 226868, in close 
proximity, orbiting about their common center of mass with a period of 5.6 days (for a review of parameters, 
see \citealt{oro2011}).  A fraction of the mass lost from the O/B star is likely accreted in the form of a 
coronal flow possibly forming an inner accretion disk, which results in either a hard or soft X-ray spectrum. The 
observations of Cygnus X-1 (hereafter Cyg X-1) over several decades record the long-lasting hard and 
soft spectral states.  Based on the extensive observational database, many theoretical investigations 
were carried out to analyze the distinct features of Cyg X-1. In particular, the spectra in the hard and soft
states and the change between spectral states have been observationally studied on the one hand, while the formation and structure
of the wind of the O/B star, including the influence of irradiation have been theoretically studied on the other.  Furthermore, 
studies of the similarities and differences in comparison with transient X-ray binaries (e.g., GX 339-4), the short-time and 
 long-time variability, and the presence of radio emission and $\gamma$-ray emission have been undertaken. 

The investigation of the structure and properties of stellar winds from massive early-type stars, as in
the case of the binary Cyg X-1, was pioneered by the work of \citet{cas1975}. The model describes the mass loss driven 
by line absorption and scattering of the supergiant radiation field (for a review see \citealt{ceh2015}). Recently,
the influence of photo-ionization caused by the feedback of X-ray radiation from the compact companion, black 
hole, or neutron star was taken into account and a new radiation hydrodynamic model was developed by \citet{ceh2015} 
and applied to the known HMXBs, especially to Cyg X-1. The three-dimensional numerical simulations provide a framework 
for the description of the wind flow between the two binary components during the two spectral states for X-ray 
luminosities of 1.9 $10^{37}$ erg/sec and 3.3 $10^{37}$ erg/sec as deduced from observations.  As the observations 
show, these two states are stable and exhibit  some scatter in the hardness and count rate. Important for the 
description is the influence of clumping, as found in recent work by \citet{kkk2018}.

Given the existence of the states characterized by (i) a low luminosity with a hard spectrum and (ii) a high luminosity 
with a soft spectrum, we examine the nature of the accretion process in the innermost region of the disk surrounding 
the black hole.  Specific attention is focused on the distribution of the mass flow in the corona and disk in this region. In a recent study  
\citet{taa2018} proposed a new accretion picture for Cyg X-1 where the condensation of gas from the 
corona onto a small or larger inner disk determines the spectrum in the hard and soft state. It was 
shown that the spectrum was dependent on the rate of mass captured from the companion star through 
the corona.  In this paper, we aim to clarify whether the spectral changes documented by the long-time 
observations validate this accretion model. 

Here we use the data of Cyg X-1 obtained with the all-sky monitor MAXI on board the Japanese module of the 
International Space Station \citep{mat2009}. The observations during the years from 2009 to 2014 have been analyzed  
in the investigation of the long-term variations in the low/hard and in the high/soft state by \citet{sug2016}. 
Our investigation is a continuation and extension of their study; we include  the data through December 2018  with the goal of focusing on the evolution of the spectral states for comparison with the new accretion model for Cyg X-1.

In Section 2 the light curves of Cyg X-1 from the MAXI observations are obtained for various energy bands to illustrate 
the repetitive phases between the hard and soft spectral states for the years between 2009 and 2018. During this 
period, the source evolved through ten episodes, each consisting of a hard spectral state followed by a soft state. In 
Section 3 the hardness-intensity diagrams (HIDs) are shown for the episodes characterized by long durations.  Since the hot 
wind from the companion star is an essential feature of the accretion model for Cyg X-1, a short review of the theoretical 
results for hot star winds, especially of the influence of irradiation on the ionization structure of the winds, is 
presented in Section 4. Of particular importance are the results of the three-dimensional time-dependent radiation 
hydrodynamic simulation of the stellar wind by \citet{ceh2015} and the influence of small-scale density inhomogeneities 
(clumping) on the strength of the wind. In Section 5 we discuss how the states of low and high luminosity (which differ by 
only a factor of two) and the spectral hard and soft states can be understood within the framework of the model of Taam et 
al. (2018) for the distribution of mass flows near the black hole.  In Section 6 the observed dipping events 
as an external cause for scatter in the hardness intensity diagrams are discussed.  Finally, we conclude in Section 7. 

\section{MAXI observations of Cyg X-1} 
\begin{figure}
  \centering
    \includegraphics [width=8.5cm]{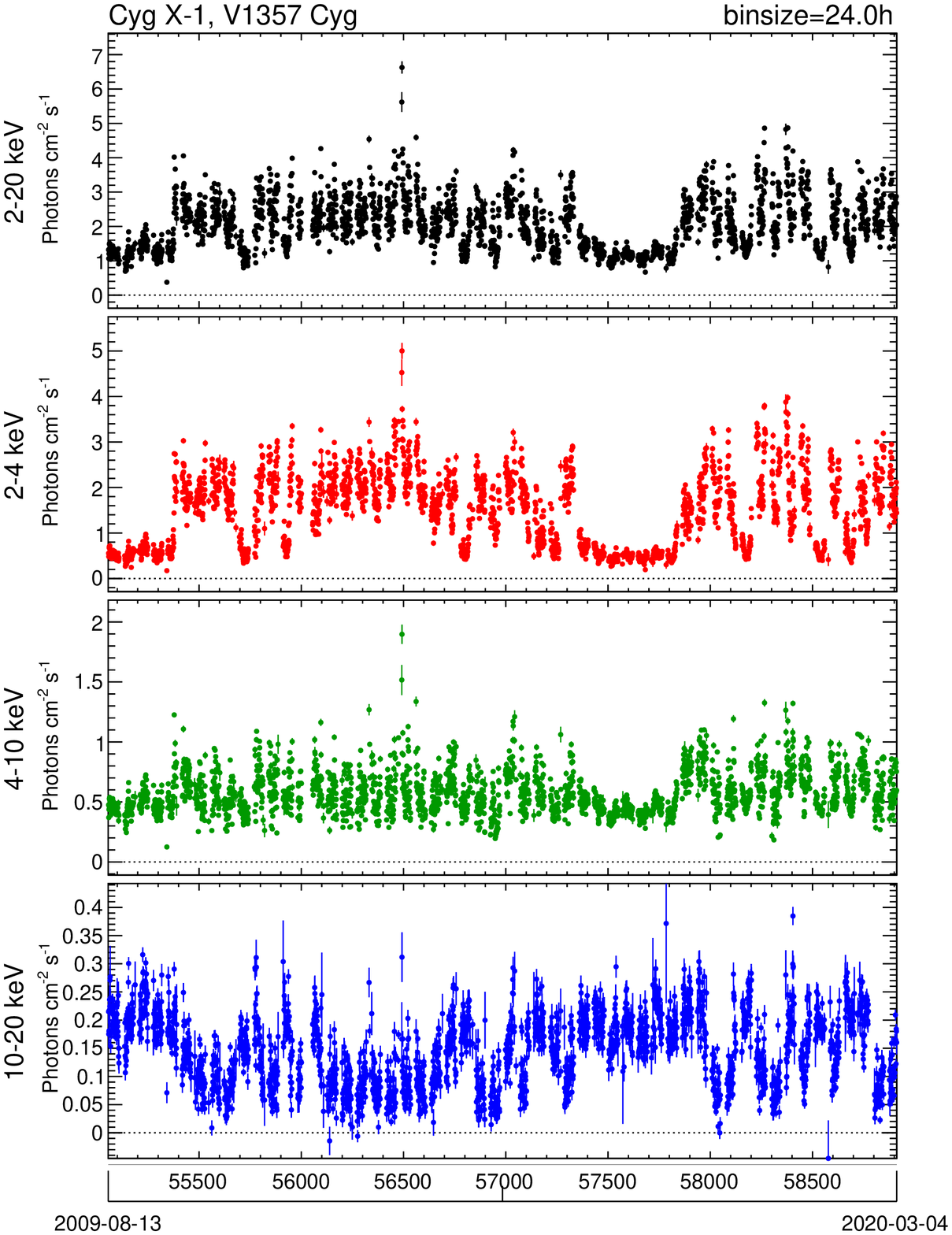}
    \caption{One-day-bin light curves of Cyg X-1 obtained with
      MAXI/GSC, processed by the latest pipeline (private communication). 
      }
    \label{f:data}
\end{figure}

Our investigation is based on observations obtained during the years from 2009 to 2018. Following the analysis of 
\citet{sug2016}, the archival one-day-bin data of Cyg X-1   can be downloaded from the MAXI home page
( http://maxi.riken.jp/top/index.php?cid=1\&jname=J1958+352).
Very recently, it was brought to our attention that 
the latest processed data for Cyg X-1, where data affected by the interference of the ISS solar panels are excluded, 
will be available on the homepage (private communication). 
With fewer zero or negative flux values in these newly processed data, the light curves are displayed for the four energy bands in Fig.\ref{f:data}.  These light curves clearly show the two different spectral states. 

 To study the HIDs the older data processed with version 5L of the data analysis package are used.
 We note that the photon counts 
with values near   or less than zero are  not 
used in the distribution of points in the HIDs.  

In \citet{sug2016} a hardness, $h$, defined as the ratio 
of counts in the energy bands  4-10 keV and 2-4 keV, and an 
intensity of counts in the energy band 2-10 keV were
used. 
The spectral states were defined as follows: high soft state (HSS) if $h<0.43$ and low hard state (LHS) if $h>0.48$. As documented by \citet{sug2016} in their Fig. 2 only 1 \%\  of the counts 
were in the intermediate hardness range between the two states during
all the years from 2009 to 2014. The variability from day to day is remarkable, and 
its analysis has been the subject of many investigations.  It should be noted that data from previous and current all sky 
monitors on RXTE, Swift-BAT, and Fermi-GBM have been also used for studies of the spectral states of Cyg X-1 \citep{gri2013}. 
A long-term  monitoring campaign of Cyg X-1 with RXTE from 1999 to 2011 \citep{gri2014} provided timing data to establish 
the relation between its timing properties and spectral states.

\section{Hardness-intensity diagrams}
The statistical analysis of the observations by \citet{gri2013} has shown that there is a high probability for Cyg X-1 
to remain in a hard or soft state. Theoretically, the results of the time-dependent hydrodynamical simulations of stellar 
wind-fed X-ray sources reveal that there are two solutions at luminosities corresponding to the hard and soft spectral state 
(for details, see the following section).  This agreement with the observations suggests that the two states are 
stable.  A graphical representation of the data in  an HID provides a clear picture of the distribution of observations 
in these two states.

\begin{figure*}
  \sidecaption  
  \includegraphics [width=10.cm]{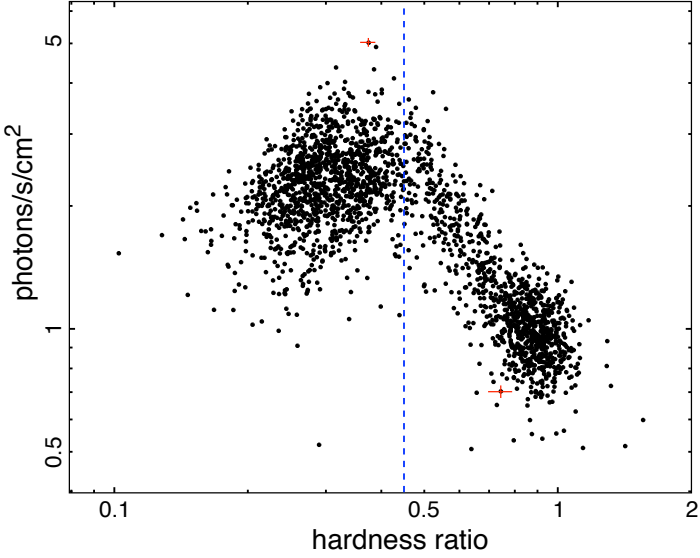}
  \caption{Hardness-intensity diagram for all MAXI observations during the time period from 2009 to 2018 for Cyg X-1.
     Photon counts in the energy band 2-10 keV are
     approximated by counts in the band 2-20 keV minus counts in 10-20 keV.
     The hardness ratio is defined as counts in the energy bands 4-10
     keV vs. 2-4 keV.  The blue vertical line indicates the threshold of
       h=0.45. Two red error bars indicate the errors at higher and lower photon count rates.
    }
    \label{f:hid}
\end{figure*}

The HID of all observations of \citet[Fig. 2]{sug2016} reveals points scattering about a hardness 
value of about  0.26 in the soft state and  0.78 in the hard state. In their Table 2, five episodes of a hard state followed by a soft state are listed.

\begin{figure*}
  \centering
  \includegraphics [width=7.cm]{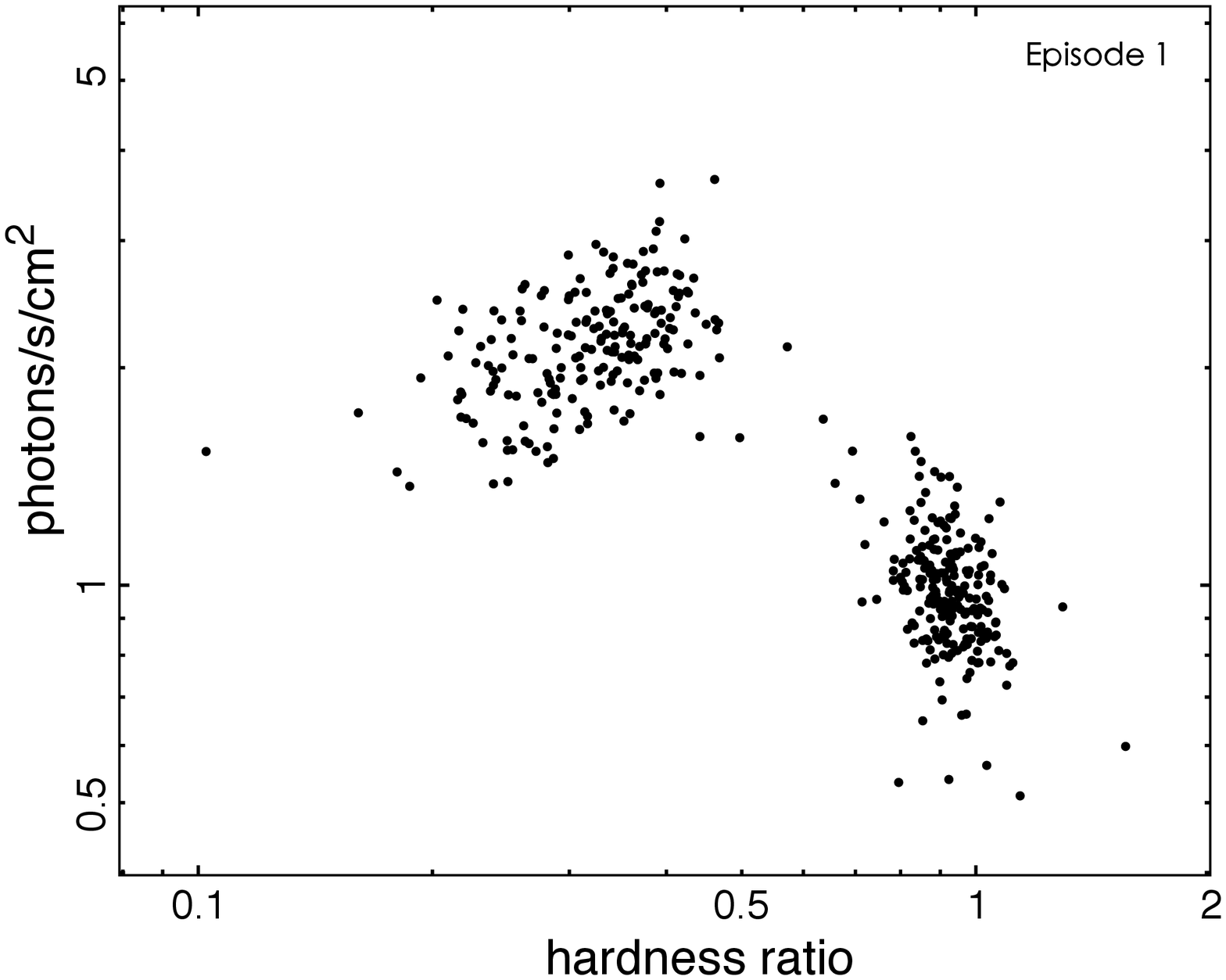}
  \includegraphics [width=7.cm]{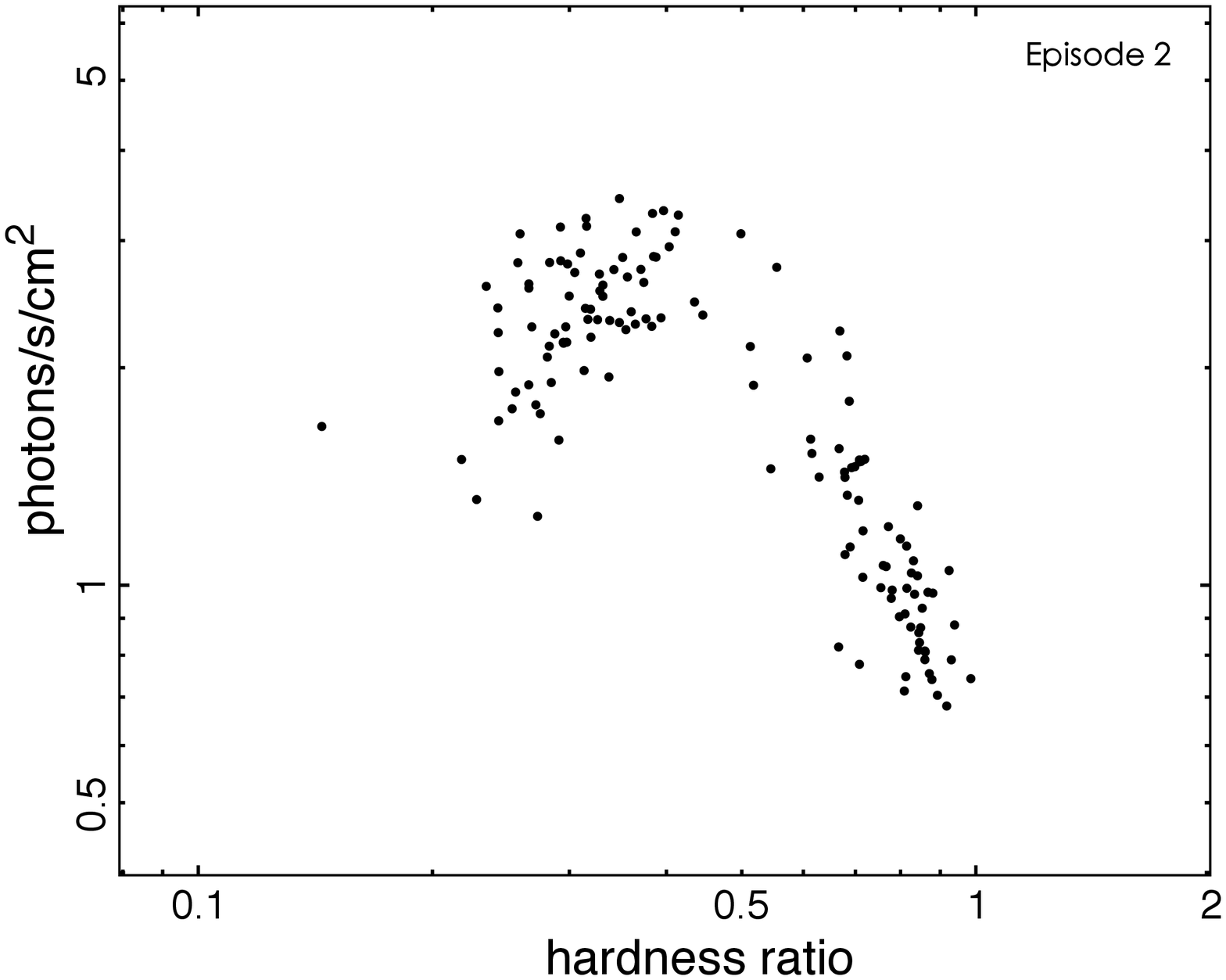}
  \includegraphics [width=7.cm]{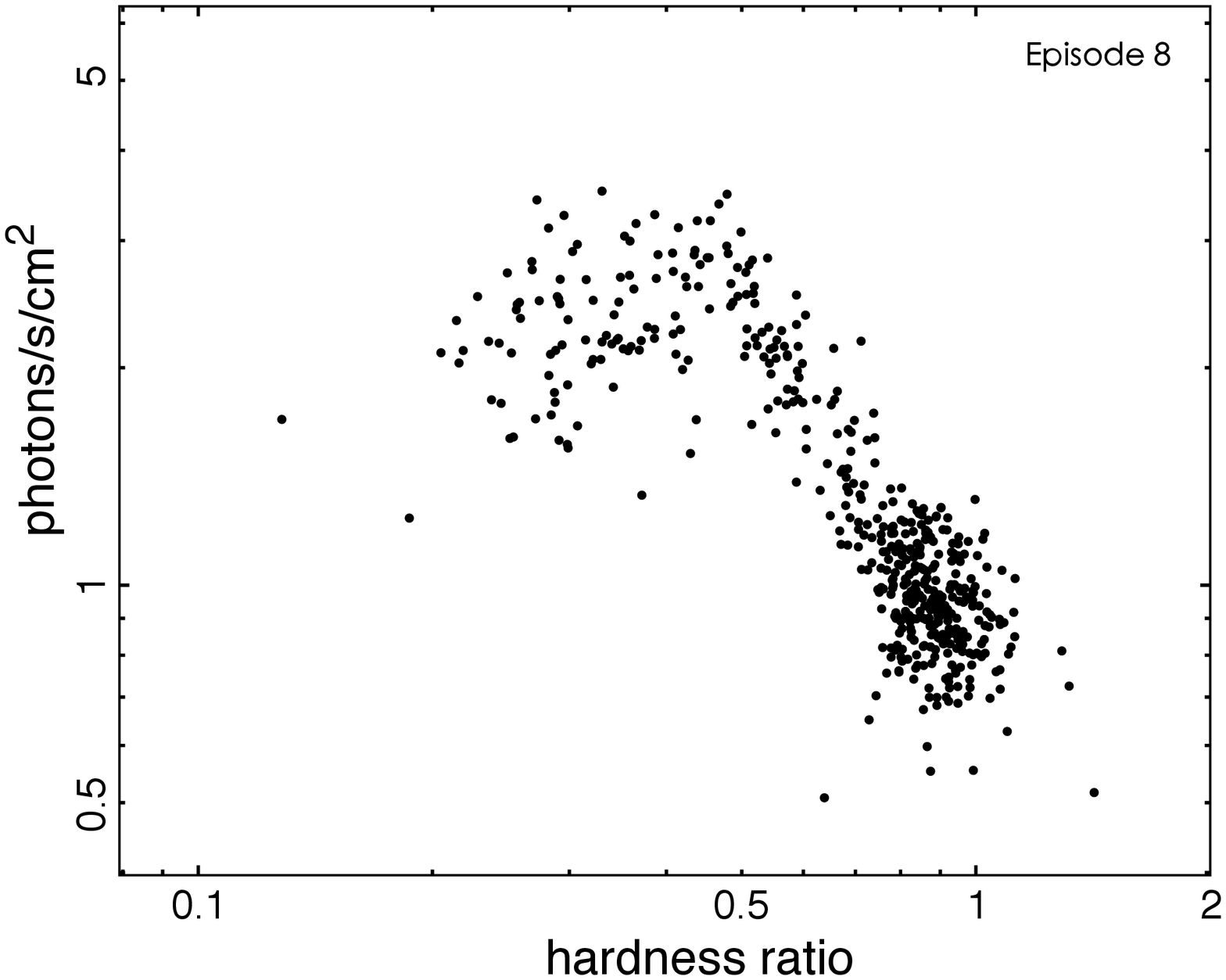}  
  \includegraphics [width=7.cm]{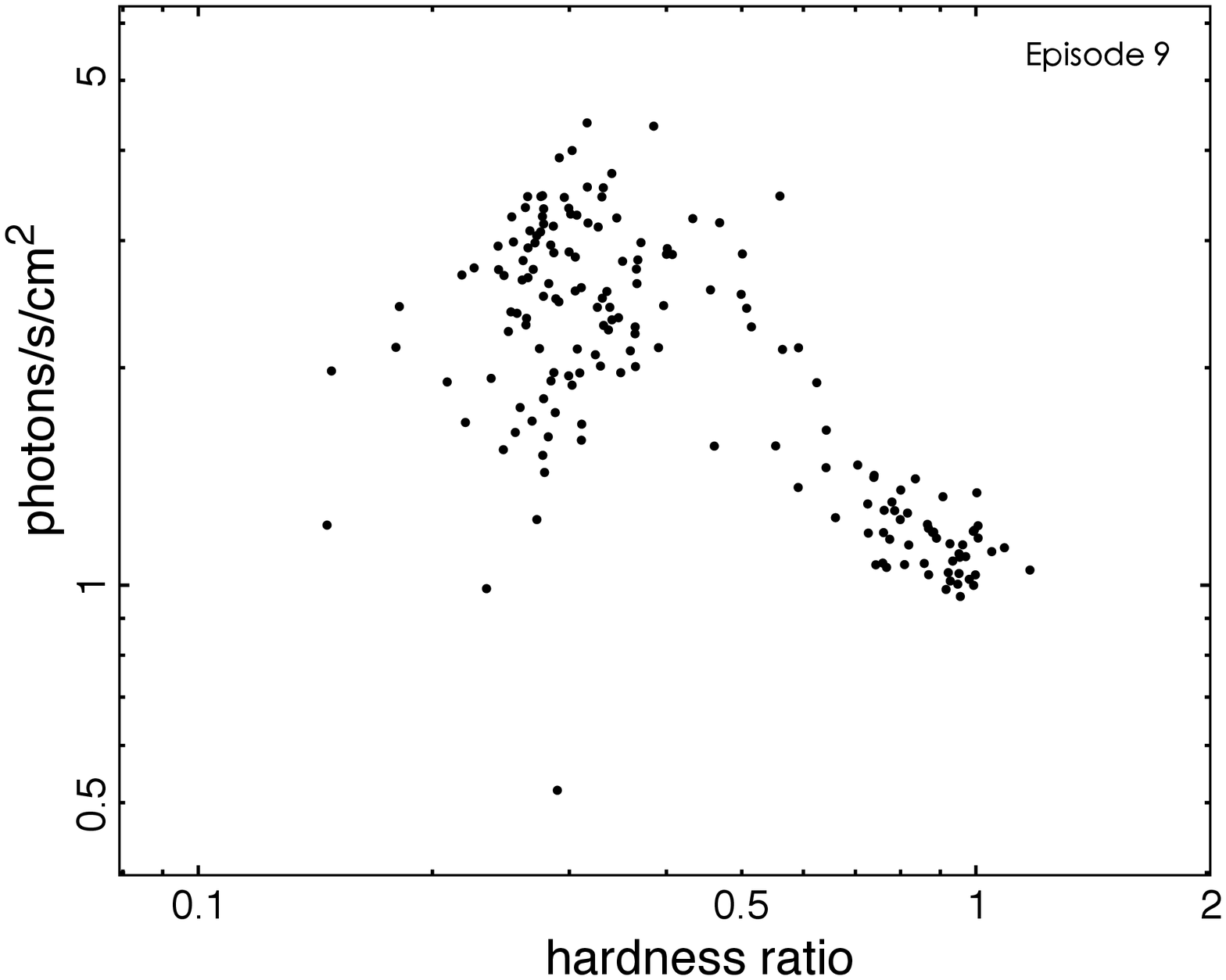}
    \caption{Hardness-intensity diagrams for   episodes 1, 2, 8, and 9
   for the hard and soft states; see the time intervals listed in Table
   1. Hardness is defined as in Fig.\ref{f:hid}.}
  \label{f:hid4}
\end{figure*}

In Fig.\ref{f:hid} we show the HID for all observations in the years
from 2009 to 2018. Since the
photon counts in the energy band 2-10 keV are not available, we used
the counts in the band 2-20 keV minus counts in 10-20 keV.
 We show representative error bars in Fig.\ref{f:hid} taken from
Fig.2 in \citet{sug2016}. The two error bars
document the larger errors at points of low count rates and smaller
errors at points of higher count rates.

During the long-term MAXI observations in which we use the hardness
 ratio of counts in the energy bands 
4-10 keV and 2-4 keV, ten episodes of a hard spectral state followed by a soft spectral state are found. 
In Table 1 the duration of the episodes is listed. These data processed using version 5L of the 
data analysis package only led to  small changes compared to the
processed data used earlier.
 We take the observations with $h<0.45$ as the soft state and with
$h>0.45$  as the hard state, adjusting the definition according to that of
  \citet{sug2016} (where they excluded the data of extremely
    rare hardness values in the range from 0.44 to 0.47 as
    intermediate state). For the long-term observations used here the
    threshold between hard and soft state is only approximate and is not as clear as in the
    data used by \citet{sug2016}. 
    This is caused by the data in
    episode 8 (see Fig.\ref{f:hid4}).

The episodes last for varying time intervals and the distribution of photon counts and hardness in 
the HID differ during the individual episodes.  The different duration
of the hard and soft state displays a random character of the transitions. In Fig.\ref{f:hid4} the HIDs for the four episodes with relatively
long durations in both spectral states are shown.

\begin{table}
\caption{Spectral states of Cygnus X-1 during the MAXI observations
  2009-2018:
data and duration of ten episodes of hard state (hardness ratio
    $h>0.45$) and soft state ($h<0.45$).
  }
\label{table:1}   
\centering       
\begin{tabular}{l l l l l l l}          
  \hline                     
  
no. & \multicolumn{2}{c}{hard state} &&  \multicolumn{2}{c}{soft state}&    \\    
& start & end  & dur.  & start & end & dur.\\
\hline 
 &(MJD)&(MJD)&(d)&(MJD)&(MJD)&(d)\\
\hline\\

1&55057 & 55378& 321& 55379& 55673& 294\\
2&55680 & 55789& 109& 55790& 55886& 96\\
3&55911 & 55941&  30& 55942& 56039& 97\\
4&56040 & 56077&  37& 56078& 56732& 654\\
5&56733 & 56746&  13& 56747& 56757& 10\\
6&56780 & 56839&  59& 56854& 57104& 250\\
7&57105 & 57268& 163& 57269& 57331& 62\\
8&57358 & 57968& 610& 57969& 58111& 142\\
9&58112 & 58201&  89& 58205& 58387& 182\\
10&58388& 58411&  23& 58441& 58480& 39\\
\hline                                            
\end{tabular}
\end{table}

\begin{figure}
  \centering
  \includegraphics[width=8.5cm]{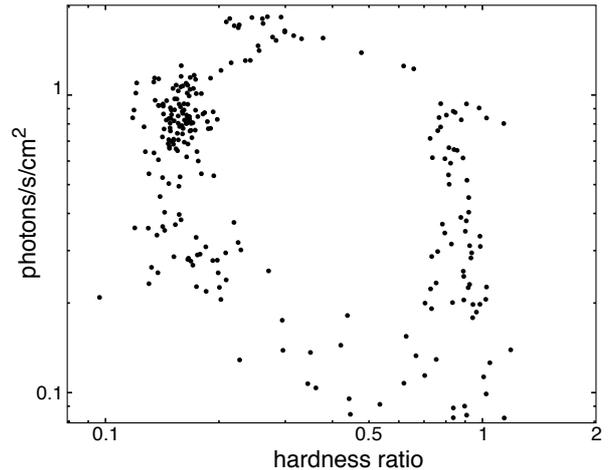}
  \caption{
        Hardness-intensity diagram for MAXI observations for GX
       339-4 during the outburst in 2010 (observations from
       December 25, 2009 to June 19, 2011. Photon counts in the energy
       band 2-10 keV; hardness: counts in the energy bands
       4-10 keV vs. 2-4 keV.}
 \label{f:hid339}
\end{figure}

The accretion process during the hard spectral state of Cyg X-1 was often compared with that in low-mass X-ray binaries (LMXBs) in the hard state,
especially GX 339-4. For comparison,  in Fig.\ref{f:hid339} we show an HID for the outburst of GX 339-4 in 2010. The data 
are based on MAXI observations from December 2009 to June 2011 (MAXI homepage, data products version L-6). The hardness 
is determined using the same energy bands as for Cyg X-1. The HID shows the data, as the source evolves, from the 
hard state of low and then increasing intensity to the transition to the soft state and the decreasing intensity
during the soft state with a return to the hard state. Upon inspection of the full cycle at both high and low 
luminosities (similar for other outbursts) the difference is clear: in LMXBs  the luminosity increase and decrease 
is much more than a factor of ten (outbursts last months) and
quiescence intervals last several years. In Cyg X-1 the 
variation in luminosity is about a factor of two and only a moderate soft state is reached.  A similar  distinction was 
found by \citet[Fig.~8]{bel2010} for RXTE/PCA observations of Cyg X-1 from 1996 to 2005 in comparison with observations 
of the 2002-2003 and 2004-2005 outbursts of GX 339-4 (where the hardness is defined as the ratio of counts in the energy 
bands 6.3-10.5 keV and 3.8-6.3 keV according to RXTE/PCA). Another striking
difference between the HIDs of LMXBs and Cyg X-1 is the appearance of
a hysteresis during the luminosity decrease which is present only for
LMXBs.

The different spectral features as shown in the HID support the view that the accretion process in Cyg X-1 differs from that 
in LMXBs. This has been studied by \citet{taa2018}, where the gas is supplied via a stellar wind, in contrast to Roche 
lobe overflow (RLOF), which plays a crucial role in the formation of a
different accretion configuration. The enhanced gas supply via a stellar
  wind leads to a perpetual existence of an inner disk and perpetual
  soft irradiation in contrast to the situation during the rise to an
  outburst in LMXBs where an ADAF fills the region near the black hole
  and hard irradiation leads to the hysteresis \citep{mhlm2005}.

\section{Properties of the stellar wind from massive early-type stars}
\subsection{Influence of X-ray irradiation on the wind in Cyg X-1}

An understanding of the soft and hard X-ray radiation emitted by Cyg X-1 as a result of the accretion of hot 
gas from the companion star requires knowledge of the properties of the stellar wind from massive early-type stars.  Models for the description of such winds were developed in the seminal work of \citet[hereafter 
CAK]{cas1975}, followed by elaborations of the theory in several subsequent investigations (for a review, see 
\citealt{ceh2015}). The mass loss from the early-type donor star occurs in form of a radiatively driven stellar wind
modulated by the tidal forces of the compact companion \citep{hace2013a}. A code for the three-dimensional
time-dependent radiation hydrodynamical simulation was further developed \citep{ceh2015}, including the
effect of the X-rays emitted from the innermost region near the black hole.  It was shown that the feedback 
effect of the ionizing radiation severely limits the efficiency of the CAK line-driven mechanism and decreases 
the radiative drag on the wind. As a result, this inhibition slows down the wind causing a broadening of 
the bow shock that forms in front of the accretion disk surrounding the compact object, thereby increasing the 
amount of material captured in the accretion disk.

Hence, the irradiation effect in Cyg X-1 leads to very different wind launching on the two hemispheres of the 
donor star. Specifically, the outflow from the hemisphere facing the X-ray source is noticeably slower than from
the opposite hemisphere lying in the X-ray shadow where the wind can be accelerated without inhibition.  This 
indicates that the X-ray irradiation leads to added accretion, and consequently to an increase in irradiation. 
However, there must be an upper limit to the irradiation since the radiative driving in the irradiated 
hemisphere is inhibited to a greater degree as the gas approaches a fully ionized state. 

\subsection{Two stable states}
In general the possible rates of wind accretion may lie in a wide range; however, for a stable steady state the  
irradiation luminosity and accretion luminosity must be consistent (for a discussion of this point, see 
\citealt[Fig.10]{kkk2018}). As shown in the results of the numerical simulations by \citet{ceh2015} two solutions for 
the mass flow in Cyg X-1 are found when feedback via irradiation is included. Specifically, solutions are found at 
a luminosity value of 3.3 $10^{37}$ erg/sec and a lower value of 1.9 $10^{37}$ erg/sec.  It is noteworthy that the observed X-ray 
luminosities in the low and high state of Cyg X-1 resemble these two values. Further support for these results is provided 
by comparison of the synthetic Doppler tomograms of theoretically predicted emission with those derived from spectra of
Cyg X-1 \citep{cevh2015}. 

The two solutions have quite a different wind flow pattern. In particular, the high-density flow in the vicinity of the 
$L_1$ point is different in the two solutions, which results in a different mass flow rate in the innermost corona, 
This is of importance for the accretion picture \citep{taa2018}, as is  discussed later. 

\subsection{Clumping and variability}

The influence of the X-rays on the wind ionization structure, addressed already in work by \citet{pau1987}, was later
reconsidered in more detailed prescriptions for the X-ray radiation feedback (for a review, see \citealt{kk2016}). 
Specifically, small-scale wind inhomogeneities \citep{kkk2018} were taken into account as a new effect. Such clumping 
in the radiation-driven winds of hot massive stars is understood to arise as a consequence of the line-deshadowing 
instability (for a review, see \citealt{pul2008}; for recent results, see \citealt{sun2018}). The clumping weakens the effect 
of the X-ray irradiation because it favors recombination.  By adopting the parameters of a binary, such a wind model can 
be used to determine the highest possible X-ray luminosity \citep{kkk2015} for wind-fed sources.   For 21 sources the 
theoretical results were compared with the observed luminosities.  The best match with the observed properties of 
HMXBs was found for models with  radially variable clumping, as suggested by \citet{naj2009}.

As noted by several authors 
(\citealt{ducci2009}, \citealt{fuerst2010}, \citealt{osk2012}), clumping, and therefore fluctuations 
in the mass accretion rate, was considered  a possible cause of the observed X-ray variability \citep{sug2016}. Since 
the effect of clumping is smaller during phases of lower irradiation \citep{kkk2018} a lower degree of variability in 
the luminosity from day to day is expected.  This can be seen in the light curves (Fig.\ref{f:data}) and is most conspicuous 
during the long-lasting low state from December 2015 to August 2017.

Based on the results of the studies incorporating the ionizing feedback regulated accretion for Cyg X-1 the following 
picture emerges:
\begin{itemize}
\item
  Two stable states exist at two distinct luminosity levels corresponding to a high luminosity for the soft
  spectral state and a low luminosity for the hard spectral state. X-ray irradiation is essential in establishing 
  these levels \citep{kkk2015}. The distribution of points in HIDs,
  especially of episode 1, suggests that rare large changes
  in irradiation can initiate a spectral state transition.
\item
  Clumping weakens the effect of irradiation, which may affect the magnetic field and the viscosity, and produces the
  X-ray luminosity variability on flow timescales  of hours to days.
\item
  The strength of clumping is a property of the wind from the O/B star and produces a limited range in the luminosity variation.
\end{itemize}

\section{Spectral properties of the two stable states originating
  from condensation?}
\subsection{Condensation in Cyg X-1}

\begin{figure*}
  \centering
  \sidecaption
  \includegraphics [width=12.cm]{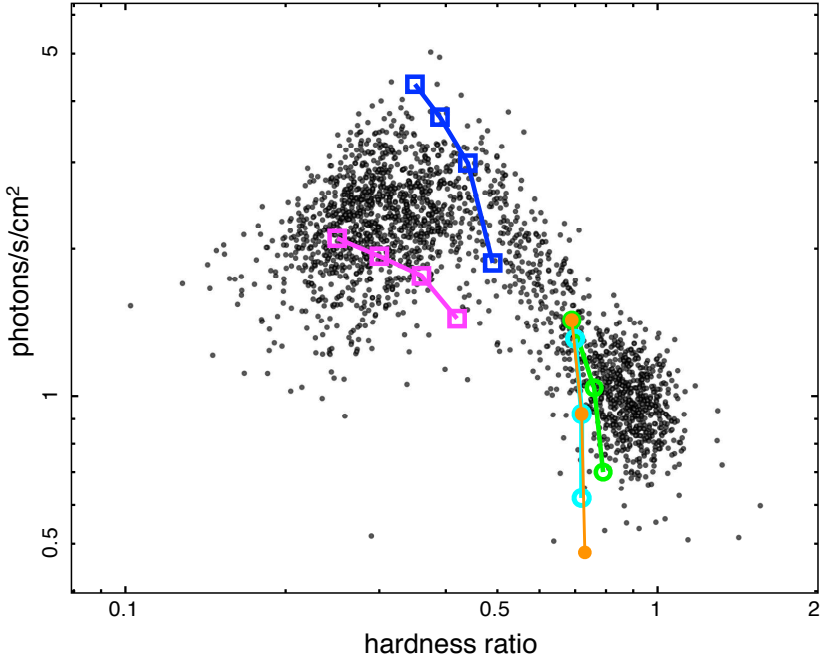}
  \caption{
      Comparison of computational
      results with MAXI observations in Fig.\ref{f:hid}.
     The hardness and count rate are computed from spectra emitted from the disk
      and corona, as determined by the condensation model.  Theoretical
      results for the soft
      state are shown for mass supply rates  $\rm{\dot{M}/\dot{M}_{Edd}}$
         =0.02,0.03,0.04,0.05 (count rates increase),
         viscosity $\alpha$=0.15 (purple dots) and
         $\alpha$=0.2 (blue dots), and albedo  fixed at 0.15. Theoretical
         results for the hard state are shown for different mass accretion
         rates ($\rm{\dot{M}/\dot{M}_{Edd}}$= 0.03, 0.035,0.04),          fixed $\alpha$ = 0.5, and albedo=0.5 (orange dots); for
         different viscosity $(\alpha=0.45,0.5,0.55)$,  a fixed
         accretion rate of 0.035, and albedo=0.5 (cyan dots); for
         different albedo ($a=0.5,0.7,0.9$),  fixed viscosity
         $\alpha=0.5$,  and an accretion rate of  0.04 (green dots). 
         }
    \label{f:hidresults}
\end{figure*}

The X-ray radiation from the innermost region of the disk, either  hard or soft, is determined by the distribution of 
the mass flow via the corona and the disk, the latter of which is caused by condensation of gas from the corona. A hard 
spectrum appears if the main accretion flow occurs in the corona as an advection-dominated accretion flow \citep{nar1995}, 
and a soft spectrum appears if the main accretion flow is in the disk.  The model for the spectral states and the transitions 
in Cyg X-1 by \citet{taa2018} describes the two states in terms of the degree to which condensation takes place, as 
determined by the strength of the wind.

The wind outflow from the companion star as described by the radiation hydrodynamic simulations by \citet{ceh2015} gives an explanation for the existence of  the two states with differing amounts of mass flow resulting from the radiatively 
driven wind either originating from the entire surface or only from the unirradiated backside of the O/B star. In this 
description, the luminosities differ by about a factor of two. In these two states not only is the amount of accretion 
different, but the structure of the wind flow is also different. 

Gas condensation from the corona to the disk is caused by  heating of the corona by the viscous dissipation releasing 
gravitational energy of the accreted gas, and cooling by vertical conduction and inverse Compton scattering of soft 
photons emitted by the underlying disk (\citet{liu2015}, \citet{qial2017}).  This condensation process was already 
discussed in several investigations for the hard state of LMXBs, where it can lead to the formation of a weak inner 
disk fed by the condensation of hot gas (\citet{liu2007}, \citet{mmh2007}, \citet{taa2008}, \citet{ldt2011}, \citet{qial2012}). 

According to the model developed by \citet{taa2018} the situation in Cyg X-1 differs from that in LMXBs.  With matter 
supplied via RLOF in LMXBs, the outer truncated disk shifts inward to the innermost stable circular orbit when the 
supply rate increases, leading to a transition from the hard to the soft state. In contrast, the increasing mass supply 
from a hot wind (only possible with wind accretion such as in Cyg X-1) results in more condensation and a stronger inner 
disk growing outward, which also leads to a transition to the soft state.  The results from the hydrodynamic 
simulations of \citet{ceh2015} reveal the increased high-density flow in the soft spectral state.

\subsection{Results of computations}
The condensation rates are calculated according to the formulae given in the investigation by \citet{ldt2011} 
and \citet{qial2012}, including the irradiation, which is important for Cyg X-1. The results for condensation yield the
mass flow rates in the corona and disk in the innermost region. For these mass flow rates spectra were calculated as 
described in \citet{qial2017}. With the spectral energy distribution from the corona and disk the hardness can 
be determined.

Our primary aim is  modeling  the radiation in the soft state because the formation of the inner disk due 
to the mass supply from the hot corona flow is the new explanation suggested by \citet{taa2018}. We computed the 
condensation of gas in Cyg X-1 for a black hole mass of 15 $\rm{M}_\odot$. For our comparison we adopt a series of 
values for the mass supply rates ranging from 0.02 to 0.05 $\rm{\dot{M}/\dot{M}_{Edd}}$ (Eddington rate
$\rm{\dot{M}_{Edd}}$=1.4$\times 10^{18}\rm{M/M_\odot g s^{-1}}$) and values for the viscosity parameter $\alpha$=0.15 
and 0.2 and a fixed albedo of 0.15 for the soft state. 

\begin{figure}
  \centering
  \includegraphics [width=8.cm]{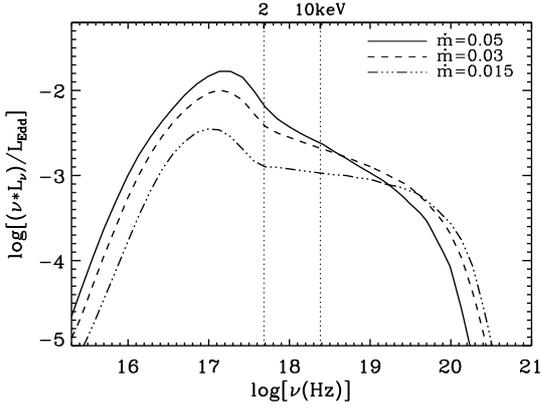}
  \caption{
    Spectra for Cyg X-1 in the soft state; condensation
  determined for a mass supply of 0.015, 0.03, 0.05
      $\rm{\dot{M}_{Edd}}$,
  $\alpha$ = 0.2, albedo=0.15 
        }
    \label{f:spectra}
\end{figure}

The relation between the count rate and hardness is shown in Fig.\ref{f:hidresults}. Typical spectra in the soft 
state are illustrated in Fig.\ref{f:spectra}.  It can be seen from the spectra that an increase in the mass supply 
rate leads to a softening of the spectrum and to an increase in the count rate in the 2-10 keV energy band.  However, 
a further increase in the mass supply rate results in a decrease in the count rate with a continuous softening of the 
spectrum as a consequence of strong condensation at a high accretion rate.  This effect is not shown in the figures 
since  Cyg X-1 does not usually  reach such a high bolometric luminosity, but it could cause the scattering in the HID 
at very low hardness and low count rates.  The two curves with slightly different viscosity values, $\alpha=0.15, 0.2$, 
indicate a sensitive dependence  on the viscosity. Therefore, the variation in the HID can be caused by a change in the
mass supply rate accompanied by a change in the viscosity in the accretion flow. 
 
\begin{figure}
  \centering
  \includegraphics [width=8.cm]{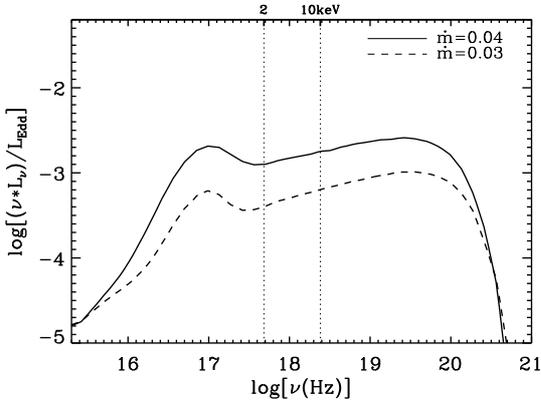}
  \caption{
    Spectra for Cyg X-1 in the hard state; condensation
  determined for a mass supply of 0.03, 0.04
      $\rm{\dot{M}_{Edd}}$,
  $\alpha$ = 0.5, albedo=0.5
        }
    \label{f:spectra-h}
\end{figure}

For the hard state, the variation in the count rate and hardness with mass supply rate is studied. In addition, the 
effects of the viscosity parameter and albedo are also explored.  Based on the luminosity in the hard state of Cyg X-1, 
values for the mass supply rate ($\rm{\dot{M}/\dot{M}_{Edd}}$ = 0.03, 0.035, 0.04) were chosen with a fixed viscosity, 
$\alpha=0.5$, and albedo, $a=0.5$, which are the same as assumed in the investigation by \citet{taa2018}. As shown in
Fig.\ref{f:hidresults}, the count rate decreases steeply with decreasing gas supply rate, while the hardness does 
not change appreciably.  The steep decrease in  the photon count rate in the 2-10 keV energy band is caused by the sudden 
decrease in soft photons from a very weak disk, which largely reduces the effect of Compton scattering. The weak disk
contributes little radiation at the energy band of 2-4 keV, leading to less variation in the hardness with accretion 
rate than that in soft state.  These properties in the hard state can be seen from the spectra in Fig. \ref{f:spectra-h}.
To check the effect of variations in the viscosity and albedo, we first fixed the accretion rate at 0.035 and the albedo at 0.5,
but varied the viscosity parameter slightly, $\alpha=0.45, 0.5, 0.55$. The results are similar to that obtained by varying 
the accretion rates. We also calculated the spectrum with $\alpha=0.5$ and $\dot M/\dot M_{\rm Edd}=0.04$ fixed while
increasing the albedo, $a=0.5, 0.7,0.9$, which yields slightly different results. The count rates increase with increasing 
mass supply, decreasing viscosity, and decreasing albedo, while the hardness only varies slightly from 0.7.       

Upon comparison of the results of the computations with the observations for the years 2009 to 2018 in 
Fig.\ref{f:hidresults}, we find that the condensation process can yield photon count rates and hardness 
in the soft state as observed for Cyg X-1 for the assumed mass supply rates, provided that the viscosity 
parameter is in the range of ~0.1-0.2. On the other hand, a higher viscosity ($\alpha\ga0.5$) and albedo ($a\ga 0.5$)
might be indicated in the hard state in addition to the contribution from possible obscuration effects (see below).

We note that the observations show a significant scatter in the data.
The  spread in the distribution of intensity-hardness values 
may be caused by the mass supply variations due to the expected
clumping in the coronal mass flow;  we note that the errors in the data
  are much smaller than this spread. This has been 
discussed by \citet{kkk2018}, who presume it to be the dominant cause of the variability. However, our results 
reveal that the condensation process strongly depends on the viscosity leading to larger variations in luminosity 
and hardness for already small viscosity changes in comparison to that for mass supply variations. It is conceivable 
that the clumpiness might affect the magnetic field structure in the disk and hence the turbulent viscosity driven 
by the action of the magneto-rotational instability. 

While the theoretical values of the intensity and hardness for the soft state span a range of the observational data 
the radiation is not as hard as observed in the hard state. A possible explanation could be due to the 
different wind flow patterns in the soft and hard state found in the hydrodynamical computations. In particular, the 
high-density flow in the proximity of the $\rm{L_1}$ point could lead to permanent absorption of the radiation from 
the weak inner disk, so that the observed radiation would appear harder than its intrinsic spectrum.  The high-density 
flow in the proximity of the $\rm{L_1}$ point in the hard state leads to so-called dipping events. However modulations of the column density are found in non-dip spectra as well (\citet{mis2016}).  We turn to 
the dipping events observed in the hard state in the next section.

\section{Dipping events in the hard state}
The scatter in the HID can arise from the dipping events associated with the strong orbital modulation of the 
column density $N_{\rm{H}}$. This modulation results in variable absorption of the X-rays 
in the stellar wind as was already found in very early work by \citet{licl1974} and which has been subsequently 
studied intensively. Very recently 
\citet{hihe2019} presented an analysis of Chandra High Energy Transmission Gratings observations of Cyg X-1
which point to the picture of a complex wind structure with temperature and density inhomogeneities. As the 
clumps pass along the line of sight, absorption dips appear in the light curve. During the canonical hard state 
strong orbital modulation of $N_{\rm{H}}$ is observed, but only very little modulation is seen during the thermally 
dominated X-ray soft state.

It is significant that the apparent difference in the degree to which clumps absorb X-ray radiation in the hard and 
soft state agrees well with the picture of the wind flow structure resulting from the radiation hydrodynamic 
simulations of \citet[Fig.9]{ceh2015}.  Specifically, a higher density stream near the black hole is found in the hard state.

We note that the new analysis of \citet{hihe2019} was taken during the low hard state. For three sections of the orbital
phase with the deepest dipping events, the observed light curve (energy band 0.5-10 keV) and measured counts in the 
0.5-1.5 keV and 3-10 keV energy bands are given with the hardness defined as the ratio of counts in the 0.5-1.5 keV 
band to the 3-10 keV energy band. During one orbital period the duration of dipping lasts for about one day (first
and  second dipping) and 1/4 day (third dipping). The daily observations affected by dipping depend on the 
particular orbital phases observed. If we adopt the results from the Chandra observations, the maximum reduction in 
the counts could be about 40\% and the hardness could be increased by a factor of two (a rough estimate because of 
the difference in the definition of the hardness).  The possible effect of dipping on the observed photon 
counts indicates that some of the data points in an HID are shifted to lower intensity and higher hardness. This 
would affect the degree to which the results from the condensation
model compare favorably with the observations 
for Cyg X-1.

\section{Conclusions}
Cyg X-1 is a particularly conspicuous high-mass X-ray binary with the transfer of matter from its blue supergiant 
companion to the black hole since  the mass transfer rate falls in a range allowing the source to be in two states, 
with either  mainly soft or hard X-ray radiation. The time dependent hydrodynamic simulations of the stellar
wind by \citet{ceh2015} have shown that X-ray irradiation of the early-type O/B companion leads to a difference in 
the stellar wind on its two hemispheres.  During the hard spectral state of Cyg X-1 the stellar wind is nearly 
symmetric, whereas during the soft state it is very asymmetric with the matter captured by the black hole mainly 
launched from the hemisphere lying in the X-ray shadow. 

Since the spectrum of the source is determined by the radiation emitted from the innermost region near the black hole, 
knowledge of the distribution of matter within the corona and the disk is necessary.  In this work, the mass flow in 
the central region depends on the interaction of the cool disk and the hot coronal gas, essentially determined by the 
condensation process of matter from the corona to the disk. It has already been shown that this hot gas condensation 
can lead to the formation of a weak inner disk in LMXBs in the hard state.

The condensation process has been applied to the formation of an inner disk in Cyg X-1 during the soft state \citep{taa2018}, 
for as strong a mass supply as possible only in wind-fed sources.  We compare the results from the condensation model 
with observations using MAXI data for the years 2009 to 2019. For this comparison we take the mass flow rates adjusted 
by condensation; determine the spectra of the radiation, theoretical values of photon count rates, and hardness; and 
make a comparison with the corresponding values from the observations presented in an HID.

As shown in Fig.\ref{f:hidresults} the data for Cyg X-1 form two clusters in the HID, one representing the hard state, 
the other representing a relatively soft state (hardness of photon counts in the energy bands 4-10 keV and 2-4 keV). 
This is different from HIDs of LMXBs where the evolution during a
single outburst forms a ``q-type'' loop  where the luminosity
increases by a factor of more than ten and a hysteresis appears
during the return of the source to the low state.

Upon comparison 
of theory with observations, we find agreement with the observations for a mass supply rate of 0.02 to 0.03 
$\rm{\dot{M}_{Edd}}$ and viscosity values of 0.15 to 0.20 for the soft state. This confirms that the inner disk can originate 
from condensation of matter from the hot coronal flow \citep{taa2018}. The scatter in the HID is not reproduced by the 
computations, but could be caused by the clumpiness of the mass flow, which can affect the magnetic field structure in 
the disk and the turbulent viscosity due to the magnetorotational instability.

For the hard state the condensation process leads to spectra that are not as hard as the observed data.  The difference 
may be attributed to the fact that the observations might be affected by absorption in the wind.  That is, dips appear 
in the light curve when clumps in the wind pass along the line of sight. In this case, the observed radiation would 
appear at higher hardness in the HID for some orbital phases. Such spectral changes during dipping were analyzed by 
\citet{hihe2019} using recent Chandra High Energy Transmission Gratings observations. It should be noted that modulations 
of the column density in spectra during non-dip events were also found (\citet{mis2016}). It would be highly desirable to 
carry out studies to examine the possible relationship between the observed high  hardness and the effect of obscuration of the 
inner disk by the wind flow in the hard state.

\begin{acknowledgements}
This research has made use of the MAXI data provided by RIKEN,JAXA
and the MAXI team.  In particular we thank Dr. Mutsumi Sugizaki for
providing new data. Financial support for this work is provided by the
National Program on Key Research and Development Project (Grant
No. 2016YFA0400804), the gravitational wave pilot B (Grant
No. XDB23040100), and the National Natural Science Foundation of China
(Grant No. 11673026,11773037).  We thank Friedrich Meyer and Weimin Yuan for discussions.\end{acknowledgements}

\bibliographystyle{aa}    
\bibliography{cyg}        

\end{document}